\pdfoutput=1
\documentclass[11pt,a4paper]{article}
\usepackage{jheppub}
\usepackage{amsthm}
\usepackage{mathtools}
\usepackage{enumitem}
\usepackage{bm}
\usepackage{hyperref}
\usepackage[capitalize]{cleveref}
\usepackage{graphicx}
\newcommand{\Q}{\mathbb{Q}}
\newcommand{\C}{\mathbb{C}}
\newcommand{\R}{\mathbb{R}}
\newcommand{\Z}{\mathbb{Z}}
\title{Varieties of four-dimensional gauge theories}
\author[a]{Ben Gripaios}
\author[b]{and Khoi {Le Nguyen Nguyen}}

\affiliation[a]{Cavendish Laboratory, University of Cambridge, J.J. Thomson Avenue, Cambridge, CB3 0HE, United Kingdom}

\affiliation[b]{DAMTP, University of Cambridge, Wilberforce Road, Cambridge, CB3 0WA, United Kingdom}

\emailAdd{gripaios@hep.phy.cam.ac.uk}
\emailAdd{kl518@cam.ac.uk}

\abstract{We use algebraic geometry to study the anomaly-free representations of an arbitrary gauge Lie algebra for 4-dimensional spacetime fermions. For irreducible representations, the problem reduces to studying the Lie algebras $\mathfrak{su}_n$ for $n\geq 3$. We show that there exist equivalence classes of such representations that are in bijection with the rational points on a projective variety that are dense in a region of the underlying real variety diffeomorphic to $\mathbb{R}^{n-3}$. It follows that the chiral ones overwhelm the non-chiral ones for $n \geq 5$. We present an efficient algorithm to find explicit anomaly-free irreducible representations and discuss the generalization to reducible representations.}
\begin{document}
	\maketitle
\flushbottom


\section{Introduction}
Our world is characterized by the fact that it is four-dimensional and has both light vector bosons and light fermions. Naturalness arguments then suggest that it is described by a gauge theory with chiral fermions, so in trying to give a more precise description (in particular one that goes beyond the Standard Model) it is thus natural to ask: which representations of a Lie algebra $\mathfrak{g}$ are both chiral and anomaly-free?

It is perhaps hard for those of us who have grown old with the 	$\overline{\mathbf{5}} \oplus \mathbf{10}$ of $\mathfrak{su}_5$ unification to appreciate how absurd it is to attempt to answer this question, {\em a priori}. For as we will see, even in the special case of representations of $\mathfrak{su}_n$ with $n \geq 3$ that are $(m-1)$-times reducible ($m\geq1$), an answer would amount to finding all whole number solutions of a polynomial equation of degree $n(n-1)/2+3$ in $m(n-1)$ variables. 

Though we expect that solutions exist for generic choices of $n$ and $m$ (because, for example, we can take chiral representations of any anomaly-free algebra containing $\mathfrak{su}_n$ and hope they restrict to chiral anomaly-free representations of $\mathfrak{su}_n$, as Georgi did with the $\mathbf{16}$ of $\mathfrak{so}_{10}$), any attempt to systematically construct them, {\em e.g.} by means of a trial-and-error scan, seems utterly hopeless: there are simply too many variables to play hit-and-hope.

Here, we introduce the use of methods of algebraic geometry to attack this problem. As we shall see, these methods turn out to be surprisingly successful in the case of irreducible representations (henceforth `irreps') of $\mathfrak{g}$ ({\em i.e.} $m=1$). Such representations, though {\em a priori} attractive to a physicist because of their minimality, turn out to be useless for the real world, because they are far too big: for $\mathfrak{su}_5$, the smallest chiral anomaly free irrep has dimension over a million. Nevertheless, there is an important qualitative conclusion, which we expect to hold for $m>1$ as well. Namely, there are a lot more chiral anomaly-free irreps than one might think on the basis of what we already know. Not only are there typically infinitely many (which is already a surprise given that there are precisely none for $\mathfrak{su}_3$ and $\mathfrak{su}_4$ \cite{Georgi_1976}) but there are as many as possible. The use of algebraic geometry allows us to make this statement precise. Namely, we recast the problem into one of finding certain rational points on a projective variety and show that these points are dense in a corresponding region of the underlying real variety. This gives hope that any reasonable attempt to answer the question we posed at the beginning in the general reducible case (even a trial-and-error scan, though we will mention some other approaches at the end) is likely to enjoy some success.

Irreducibility results in a big simplification because it allows us to restrict our attention to the cases where $\mathfrak{g} = \mathfrak{su}_n$, with $n\geq 3$. The argument is as follows. 
There are no non-trivial anomaly-free irreps if $\mathfrak{g}$ has a non-trivial abelian summand, so we may take $\mathfrak{g}$ to be semisimple without loss of generality; for semisimple $\mathfrak{g}$, the anomaly cancels if and only if it cancels for each simple summand, so we may take $\mathfrak{g}$ to be simple without loss of generality; for simple $\mathfrak{g}$, every representation is anomaly-free, unless $\mathfrak{g} = \mathfrak{su}_n$, with $n\geq 3$.

The state-of-the-art regarding our understanding of chiral anomaly-free irreps of $\mathfrak{su}_n$ can be described in few words. We have already seen that there are none for $n=3$ and $n=4$, while a trial-and-error scan \cite{Eichten_1982} shows the existence of at least a handful (at most 8) for $n \in \{5,6,7,8,10,12,14,16\}$; beyond that, it is a case of {\em hic sunt leones}.

From this scant knowledge, one might easily infer that such irreps are somewhat few-and-far-between. As we shall see here, the opposite is true: there are infinitely many such irreps of $\mathfrak{su}_n$ for each $n \geq 5$ (as usual, in this counting we identify representations that differ by a unitary transformation); moreover, they utterly overwhelm the {\em non}-chiral anomaly-free irreps (in a sense that we will make precise below).\footnote{We expect that the same is true for reducible representations.}

Since there are infinitely many such irreps, their characterization becomes a delicate matter. We do it by organizing the irreps into equivalence classes (each of which contains infinitely many irreps), on which the notions of being chiral and anomaly-free are well-defined. Doing so allows us to make contact with algebraic geometry, because the
anomaly-free classes are in bijection with a subset of the rational points on a projective variety (namely a cubic hypersurface). This variety is very special: not only is it a {\em rational}\footnote{We embrace the algebraic geometers' convention of using the word `rational' to mean (at least) two different things. Here `rational variety' means, roughly speaking, that the variety can be parameterized using rational functions (more precisely, it is birationally equivalent to projective space); earlier, `rational point' meant a point in a variety over $\Q$.} variety (over $\C$, $\R$, or $\Q$), but it also has a large group of automorphisms (namely the permutation group on $n$ objects). Together, these properties enable us to give a complete description of the chiral anomaly-free irreducible representations.

In particular, we are able to characterize the 
rational points corresponding to anomaly-free irreps (both chiral and non-chiral) by studying the underlying real variety. To wit, we show that these points are dense in a region of the real variety that is diffeomorphic to $\R^{n-3}$. This simple result is somewhat surprising given that the real variety is topologically rather complicated (for $n=5$, for example, it is diffeomorphic to the connect sum of seven copies of $\R P^2$), and indeed it will take us some effort to prove.  
It is much easier to prove that the subset of these rational points corresponding to non-chiral anomaly-free irreps  are dense in a subregion that is diffeomorphic to $\R^{(n-3)/2}$ for odd $n$, or $\R^{(n-2)/2}$ for even $n$; put together, these results give precise meaning
to our earlier claim that the chiral anomaly-free irreps overwhelm the non-chiral ones for all $n \geq 5$.

Now that we have a complete topological picture in hand, we may turn to the actual business of parameterizing the anomaly-free irreps. In principle this is straightforward, because of the fact that our variety is rational over $\Q$. Explicitly, we can employ a generalization of the method of secants (developed to find the anomaly-free representations of $\mathfrak{g}=\mathfrak{u}(1)$ in Ref. \cite{Allanach_2020}). A na\"{\i}ve application of this results in a rather inefficient algorithm, because most of the time it produces rational points lying outside of the region of the variety that corresponds to representations. But now the automorphism group of the variety comes to our rescue. Indeed, the orbit under the action of $S_n$ on the region of interest is dense in the variety. 
Using this fact, one can produce an algorithm that almost always outputs anomaly-free irreps.  

At least for $n = 5$, this ‘almost always’ is much stronger than one might na\"{\i}vely imagine, because there are only a finite number of rational points (25 in fact) on the variety that are `bad', in the sense that they do not correspond to anomaly-free irreps up to a permutation. This is surprising because the corresponding real points form a union of 1-dimensional smooth manifolds, so are infinite in number.	To show this requires us to use the theory of elliptic curves (namely, we compute the associated Mordell-Weil groups), so we do not know how to generalize it to $n>5$. It is nevertheless tempting to conjecture that the bad rational points, (unlike the `good' ones) satisfy at least the weak property of not being dense in their corresponding real points. The upshot, at least for $n=5$, is that there are very few bad points for our algorithm to hit, and so it almost never fails to come up with the goods.

\begin{figure}
	\centering
	\includegraphics[width=\linewidth]{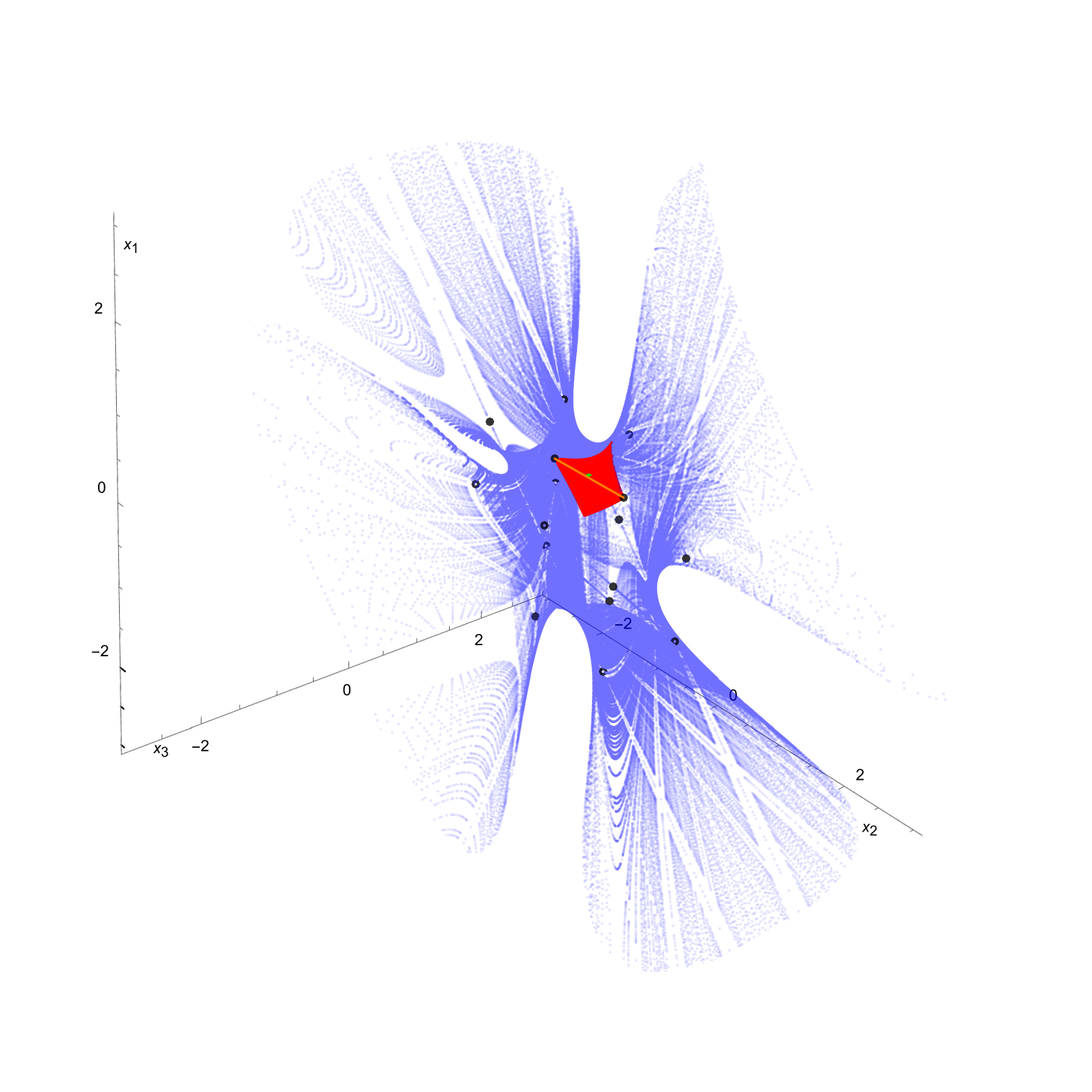}
	\caption{Rational points on an affine patch of the Clebsch diagonal surface, with coordinates $x_{1,2,3}$ as defined in \cref{eq:xi}. Each red point corresponds to an infinite class of chiral anomaly-free irreps of $\mathfrak{su}_5$. The points in orange correspond to non-chiral irreps. The blue points are rational points in the $S_5$ orbits of red ones, so can be used in our algorithm to find anomaly-free irreps, while the black points are not in such orbits, and cannot be used. The green point corresponds to the trivial representation.}
	\label{fig:figclebschscanv7}
\end{figure}

\begin{figure}
	\centering
	\includegraphics[width=\linewidth]{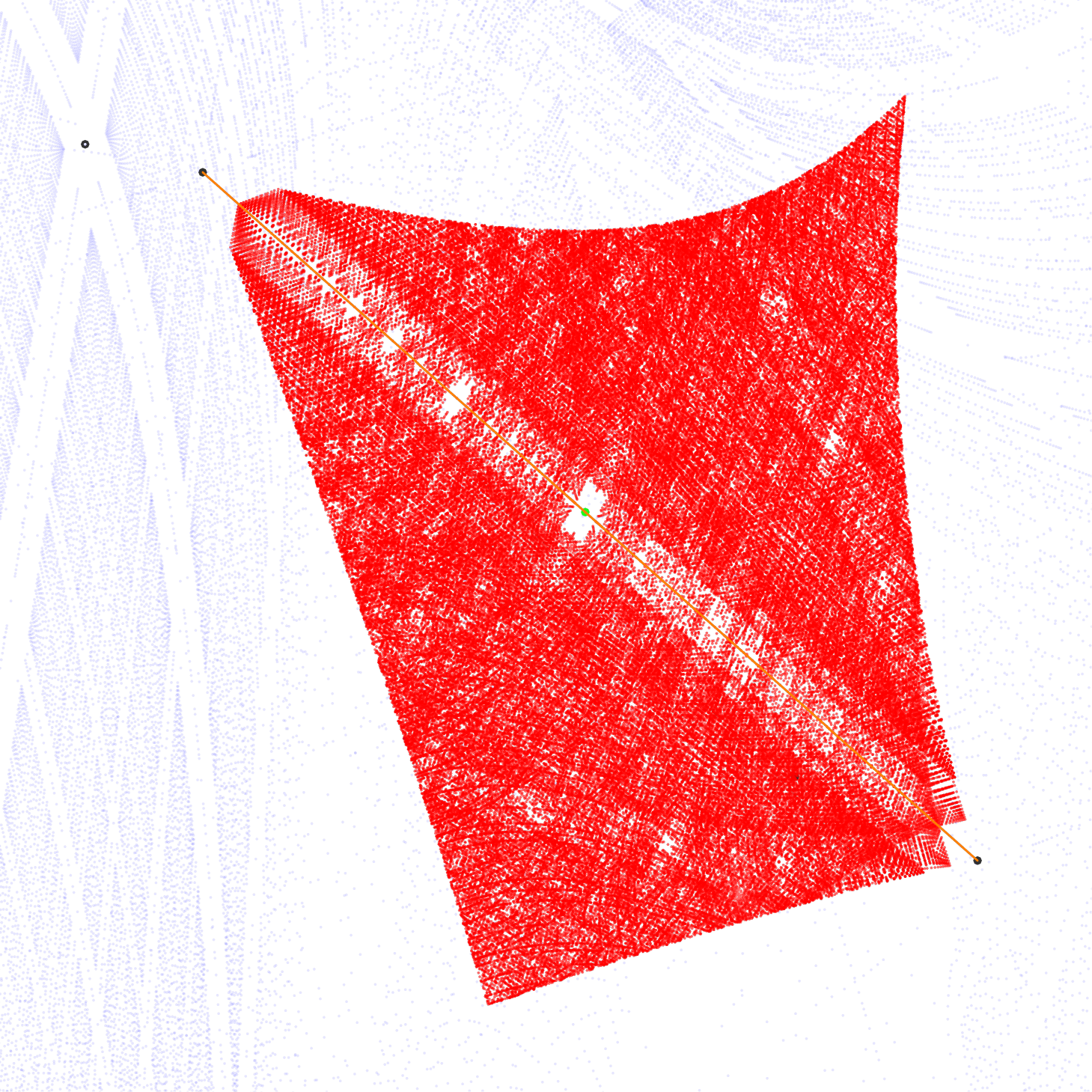}
	\caption{Close-up view of the region of interest in \cref{fig:figclebschscanv7}.}
	\label{fig:figorderedregionzoomedinv6}
\end{figure}

To illustrate all of this, we show in \cref{fig:figclebschscanv7,fig:figorderedregionzoomedinv6} the results for $n=5$ obtained using our algorithm on a modern portable computer. The variety in this case is none other than the Clebsch diagonal surface, perhaps the most celebrated surface in mathematics. The algorithm took just 480s to find the $117\, 143$ rational points shown in red (each of which, we recall, corresponds to infinitely many chiral anomaly-free irreps), along with the non-chiral anomaly-free irreps shown in orange.
This is to be compared with a brute-force trial-and-error method (as in Ref. \cite{Eichten_1982}), which took 650s to find just 60 points.
 
As expected, the red points appear to be dense in a region diffeomorphic to $\R^2$. The blue points show rational points that are in the orbit of the red region under permutations. They appear to be dense in the underlying real Clebsch diagonal surface and indeed they do a good job of illustrating the latter's topology. The black points are the aforementioned bad points (of which only 16 out of 25 are visible on the affine patch shown in \cref{fig:figclebschscanv7}). 

With our new understanding of the chiral anomaly-free irreps, we may turn to physics. The bad news, is that the smallest chiral anomaly-free irrep of $\mathfrak{su}_5$, in terms of dimension, has dimension $1\,357\,824$. So the experimentalists will need to find many more particles before the phenomenologists show an interest in our results. We do know, however, that just generalizing to {\em once}-reducible representations, we can get anomaly-free chiral representations whose dimensions are not only much lower, but are also of great phenomenological interest, {\em e.g.} $\overline{\mathbf{5}} \oplus \mathbf{10}$. Thus we turn to the reducible case. 

For general $\mathfrak{g}$, there seems to be very little one can do here: the variety that results is just too icky. But if we restrict to semisimple $\mathfrak{g}$, we again find a single homogeneous polynomial equation, so we can once more play the trick of producing many chiral anomaly-free representations given just one (for example, starting from the $\overline{\mathbf{5}} \oplus \mathbf{10}$ anomaly-free representation of $\mathfrak{su}_5$, we obtain $\mathbf{10240}\oplus\overline{\mathbf{5120}}$ by doubling, and so on). However, our polynomial is now not a cubic, but rather has degree $n(n -1)/2 + 3$. As a result, the basis of the method of secants -- that through every two rational points on a cubic there exists a line and this line generically intersects the cubic in a third rational point -- fails.\footnote{One hope here would be that these polynomials have singularities of the maximum possible degree, so that we can use the method of tangents. Alas, they do not.}

We can go much further if we consider not arbitrary reducible representations of a Lie algebra $\mathfrak{g}$, but rather representations that are built as products of irreps. Sadly, not every representation can be built in this way ({\em e.g.} $\overline{\mathbf{5}} \oplus \mathbf{10}$ of $\mathfrak{su}_5$), but some of the representations that result have much smaller dimensions, so perhaps might be of use for phenomenology. We will report on this elsewhere.

A further trick, which we do not explore further here, is that having found chiral anomaly-free representations of either $\mathfrak{su}_{k+l}$ or $\mathfrak{su}_{kl}$, we can restrict to their obvious respective $\mathfrak{su}_{k}$ subalgebras to obtain reducible (in general) chiral anomaly-free representations. By stripping off non-chiral summands, one again obtains smaller, chiral, anomaly-free representations.

\section{Irreps of $\mathfrak{su}_n$ and anomaly cancellation \label{sec:sigma}}
As is well-known, Weyl's theorem of the highest weight \cite{Humphreys_1972} tells us that the irreps of $\mathfrak{su}_n$ are in bijection with the
$(n-1)$-tuples $(m_1,\dots,m_{n-1})$ of non-negative integers.\footnote{In the language of Young tableaux, $m_i$ is number of columns with $i$ boxes.} The irrep dual to $(m_1,\dots,m_{n-1})$ is given by $(m_{n-1},\dots,m_{1})$; we say that an irrep is {\em non-chiral} if it is self-dual and {\em chiral} otherwise.

For our purposes, it is much more convenient to shift the $m_i$ up by one, defining the $(n-1)$-tuple of positive integers $q=(q_1,\dots,q_{n-1}):=(m_1+1,\dots,m_{n-1}+1)$. Indeed, in terms of the $q_i$, the anomaly cancels if and only if \cite{Georgi_1976} the homogeneous cubic polynomial
\begin{equation}
\sum_{i,j,k=1}^{n-1}a_{ijk}q_iq_jq_k\label{eqn:anom_irrep}
\end{equation}
vanishes, where $a_{ijk}$ is completely symmetric in $i$, $j$, and $k$ and is given for $i\leq j\leq k$ by 
\begin{equation}
	a_{ijk}=i(n-2j)(n-k). \label{eqn:aijk}
\end{equation}
Since ${a_{ijk}=-a_{n-i,n-j,n-k}}$, one checks that an irrep is anomaly-free iff its dual is and that every non-chiral irrep is anomaly-free.

Our first observation is that the homogeneity of the polynomial in $q_i$ implies that $(q_1,\dots,q_{n-1})$ is anomaly-free or chiral iff $(\lambda q_1,\dots,\lambda q_{n-1})$ has the same property for every positive integer $\lambda$. So given one irrep with such a property, one can immediately find infinitely-many others with the same property.

To make further progress, it is 
helpful to carry out yet another change of co-ordinates. Following Okubo \cite{Okubo_1977}, we introduce co-ordinates $\sigma_i$ with $i \in \{1,\dots,n\}$ defined by\footnote{Our $\sigma_i$ are $n$ times Okubo's; defining them in this way means they take values in the integers.}
\begin{equation}
\sigma_i:=\left(-\sum_{k=1}^{i-1}kq_k+\sum_{k=i}^{n-1}(n-k)q_k\right).\label{eqn:sigma_from_q}
\end{equation}
We have gone from $n-1$ co-ordinates $q_k$ to $n$ co-ordinates $\sigma_i$. Correspondingly, we have that
\begin{equation}
	\sum_{i=1}^{n}\sigma_i=0. \label{eqn:sigma1}
\end{equation}
These co-ordinates take values in the integers and using
\begin{equation}
	\sigma_i-\sigma_{i+1}=nq_i,\,i\in\{1,\dots n-1\}, \label{eqn:q_from_sigma}
\end{equation}
we see that the adjacent $\sigma_i$ are constrained to differ by positive multiples of $n$.

The beauty of Okubo's $\sigma_i$ (at least in the eyes of these beholders) is that they allow us to see and exploit the symmetry of the anomaly cancellation condition following from \cref{eqn:anom_irrep}, which now reads
\begin{equation}
  \sum_{i=1}^{n}\sigma^3_i = 0.	\label{eqn:sigma3}
\end{equation}
      This, together with \cref{eqn:sigma1}, is invariant under permutations of the $\sigma_i$. As a result, given any solution of these two equations in the integers, we can permute them to find another solution. In particular, we can always find a solution with $\sigma_i\geq \sigma_{i+1}$, for $i\in\{1,\dots,n-1\}$, by reordering. Now, if these inequalities happen to be strict, {\em i.e.} if no two $\sigma_i$ are equal, then by multiplying by a sufficiently large positive integer, we can find a solution satisfying \cref{eqn:q_from_sigma} for some positive $q_i$ and hence corresponding to a {\em kosher} anomaly-free irrep. Thus almost all integer solutions of equations \cref{eqn:sigma1} and \cref{eqn:sigma3} correspond to anomaly-free irreps. This idea will be key to the design of an efficient algorithm for finding such irreps.

      There is another remarkable feature of Okubo's reformulation that deserves comment. The integer solutions of \cref{eqn:sigma1,eqn:sigma3}, without further restriction on the $\sigma_i$, are precisely the anomaly cancellation conditions for an $(n-1)$-times reducible representation of $\mathfrak{g} =\mathfrak{u}_1$ (i.e. electromagnetism with $n$ charged chiral fermions). This coincidence seems to us to be nothing short of miraculous, not least because the $\sigma_i$ in the $\mathfrak{g} =\mathfrak{su}_n$ cannot be interpreted as the charges obtained by restricting an irrep to some $\mathfrak{u}_1$ subalgebra of $\mathfrak{su}_n$. Whatever the explanation of this miracle may be, it does suggest a way of trying to solve the equations. Namely, we use methods of projective geometry, just as was done in the $\mathfrak{u}_1$ case in Ref. \cite{Allanach_2020}. 

\section{The geometric picture \label{sec:dio}}
In the last Section, we recast the problem of finding the anomaly-free irreps of $\mathfrak{su}_n$ into a problem of finding solutions to homogeneous polynomial equations taking values in subsets of the integers (namely, we require the $q_i$ to be positive, while we require that consecutive $\sigma_i$ differ by a positive multiple of $n$). In this Section, we recast this diophantine problem into one of projective geometry.

To do so, we consider the projective space $\Bbbk P^{n-1}$, with homogeneous co-ordinates $[\sigma_1:\dots:\sigma_n]$. It is convenient to let the field $\Bbbk$ be any one of $\C,\R,$ or $\Q$. A point in $\Bbbk P^{n-1}$ is thus given by an equivalence class of $(\sigma_1,\dots,\sigma_n) \in \Bbbk^n$, where the equivalence relation is given by $(\sigma_1,\dots,\sigma_n) \sim (\lambda \sigma_1,\dots, \lambda \sigma_n)$ for any non-zero $\lambda \in \Bbbk$. The homogeneous polynomials appearing in 
\cref{eqn:sigma1,eqn:sigma3}, namely $\sum_{i=1}^n \sigma_i$ and $\sum_{i=1}^n \sigma_i^3$, then define a projective variety $V_n$ in  $\Bbbk P^{n-1}$, which will be the main protagonist in our story. The polynomial $\sum_{i=1}^n \sigma_i$ on its own simply defines a projective subspace of  $\Bbbk P^{n-1}$ which we simply denote by $\Bbbk P^{n-2}$ and for which we use homogeneous co-ordinates $[\sigma_1:\dots:\sigma_{n-1}]$ (so we have $\sigma_n = 	-\sum_{i=1}^{n-1}\sigma_i$ in $\Bbbk P^{n-2}$).

To deal with the additional restrictions on the integer solutions to \cref{eqn:sigma1,eqn:sigma3} that correspond to anomaly-free irreps, it will be convenient to single out a number of subsets of $\Bbbk P^{n-1}$. These are not, in general, varieties, so we call them {\em regions} instead. To wit, we define:
\begin{enumerate}
\item The  {\em unorderable} region, for which at least two $\sigma_i$ coincide, together with its complement, the {\em orderable} region;
\item For $\Bbbk \neq \C$, the {\em weakly-ordered} region, for which either $\sigma_1 \geq \sigma_2 \geq \dots \geq \sigma_n$ or $\sigma_1 \leq \sigma_2 \leq \dots \leq \sigma_n$;
\item For $\Bbbk \neq \C$, the {\em ordered} region, in which we replace the weak inequalities in the previous definition by strict ones.
\end{enumerate}
We observe that the action induced on the ordered region by the action of $S_n$ on $\sigma_i$ is stabilized by a $\Z/2$ subgroup whose non-trivial element reverses the ordering of the $\sigma_i$. From here, it is not hard to deduce that the orderable region is made up of $n!/2$ disjoint copies of the ordered region.

Similarly, in order to distinguish in the geometric picture between chiral and non-chiral irreps, we define:
\begin{enumerate}[resume]
\item the  {\em palindromic} region, for which $[\sigma_1:\dots:\sigma_{n}] = [-\sigma_{n} : \dots : -\sigma_1]$, together with its complement, the {\em nonpalindromic} region.
\end{enumerate}

From here, we define the corresponding regions {\em on a variety} in $\Bbbk P^{n-1}$ (which could, for example, be either $V_n$ or $\Bbbk P^{n-2}$) as the intersections of the above regions with the variety.

To give an inkling of the rationale behind these definitions, the reader may wish to check that the nonpalindromic ordered rational points on $V_n$ correspond, after clearing denominators, to the chiral anomaly-free irreps that are our heart's desire.

For the purpose of making calculations (and drawing pictures), it is useful to note that the ordered region on $\Bbbk P^{n-2}$ is contained in the affine patch defined by $\sigma_n \neq 0$ (for, if  $\sigma_n$ were to vanish in the ordered region, then all of the other $\sigma_i$ would have to be of the same sign, so could not sum to zero). For this patch, we use co-ordinates
\begin{equation} \label{eq:xi}
   (x_1,\dots,x_{n-2}):=\left(-\frac{\sigma_1}{\sigma_n},\dots,-\frac{\sigma_{n-2}}{\sigma_n}\right),
 \end{equation}
such that $-\frac{\sigma_{n-1}}{\sigma_n}=1-x_1-\dots-x_{n-2}$. 

As $\Bbbk\neq\C$ in the definition of the (weakly) ordered region, it will often be convenient to insist on using homogeneous co-ordinate representatives in which $\sigma_1$ is non-negative. It follows that $\sigma_1$ is positive in the ordered region on $\Bbbk P^{n-2}$ and that $\sigma_n$ is negative.

For example, the trivial representation has highest weight $(m_1,\dots,m_{n-1})=(0,\dots,0)$ and corresponds to a point with the various coordinates $(q_1,\dots,q_{n-1})=(1,\dots,1)$, \linebreak $[\sigma_1:\sigma_2:\dots:\sigma_{n-1}:\sigma_n]=[n(n-1)/2:n(n-3)/2:\dots:-n(n-3)/2:-n(n-1)/2]$ and $(x_1,x_2,\dots,x_{n-2})=(1,(n-3)/(n-1),\dots,-(n-5)/(n-1))$.

Our goal now is, for each $n$, to characterize the nonpalindromic ordered rational points on $V_n$ as fully as possible. The main results in that direction will be firstly that the ordered region on $V_n$ over the reals is diffeomorphic to $\R^{n-3}$ and that the rational points are dense in it and secondly that one can devise an efficient algorithm to find those rational points. 
\section{The case of $\mathfrak{su}_5$}
To ease the reader's passage, we begin with a discussion of the case $n=5$. Not only is this case the first non-trivial one (readers are invited to work out the details for $n=3$ and $n=4$ for themselves), but also it allows us to draw pretty pictures, as in \cref{fig:figclebschscanv7,fig:figorderedregionzoomedinv6,fig:clebschwithlines,fig:figclebschellipticcurvesv6,fig:figorderedregionv2}. Happily, most of the arguments we present for $n=5$ generalize straightforwardly to any odd $n$ and we will do so in the next Section. For even $n$ there are some differences, for which the reader will need to be on their guard.
\subsection{$V_5$ is a rational variety}\label{sec:rational}
The variety $V_5$ is the famous Clebsch diagonal cubic surface (see {\em e.g.} Refs. \cite{Huybrechts_2023,Segre_1942}). Crucial to our story is the fact that it contains many lines. Indeed it is smooth, so just like any other cubic surface it must contain
27 lines over $\C$. But it is special among smooth cubic surfaces in that it also contains 27 lines over $\R$. Fifteen of these lines are given by
\begin{equation}
	\sigma_i=\sigma_j+\sigma_k=\sigma_l+\sigma_m=0,\label{eqn:rational_lines}
\end{equation}
with $\{i,j,k,l,m\}=\{1,2,3,4,5\}$, while the remaining twelve are given by
\begin{equation}
	\sigma_i+\varphi\sigma_j+\sigma_k=\varphi\sigma_i+\sigma_j+\sigma_l=-\varphi(\sigma_i+\sigma_j)+\sigma_5=0,
\end{equation}
with $1\leq i<j\leq 4$ and $\{i,j,k,l\}=\{1,2,3,4\}$, and $\varphi=\frac{1+\sqrt{5}}{2}$ is the irrational golden ratio. It thus follows that $V_5$ has 15 lines over $\Q$. The rational line with $\sigma_3 = \sigma_1+\sigma_5=\sigma_2+\sigma_4$ is the palindromic region. 

\begin{figure}
	\centering
	\includegraphics[width=\linewidth]{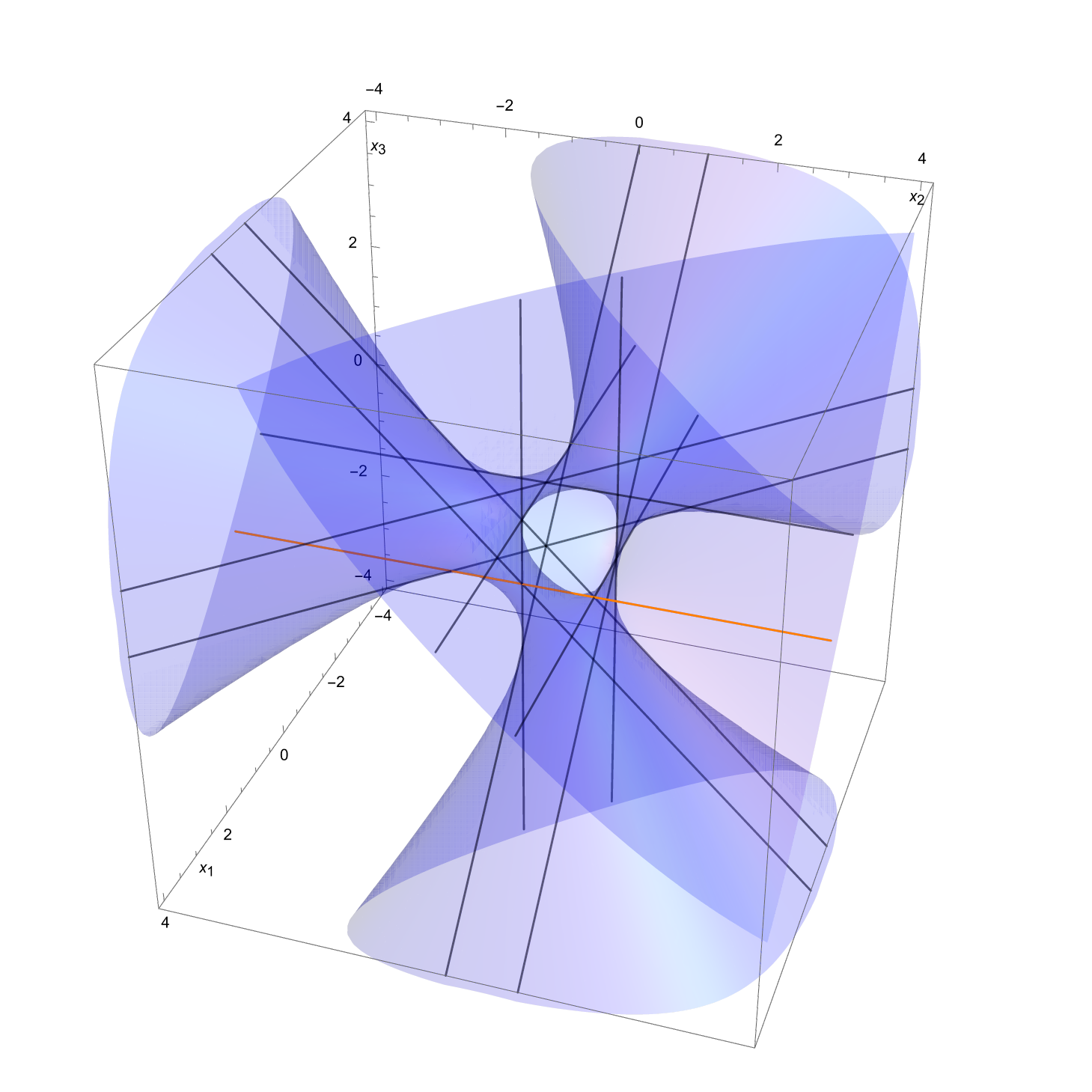}
	\caption{An affine patch of the Clebsch diagonal surface (in blue), along with the palindromic region (in orange) and the remaining 11 rational lines (in black). The co-ordinates $x_{1,2,3}$ are defined in \cref{eq:xi}.}
	\label{fig:clebschwithlines}
      \end{figure}

Fig.~\ref{fig:clebschwithlines} shows the affine patch of the real variety with $\sigma_5 \neq 0$, along with the 12 rational lines that intersect it (one of which is the palindromic region).

      The importance of the rational lines are that they enable us to show that $V_5$ is a rational variety over $\Q$ (as well as over $\R$ and $\C$). This is turn allows us to parameterize the surface in a simple way and to find it its rational points. 

      The proof that $V_5$ is a rational variety over $\Q$ hinges on the fact that there exists a pair of disjoint rational lines in $V_5$ (for example the pair of lines with $\sigma_1=\sigma_2+\sigma_4=\sigma_3+\sigma_5=0$ and $\sigma_1+\sigma_4=\sigma_2+\sigma_5=\sigma_3=0$). Indeed, given any two distinct rational lines $\Gamma_{1,2}$ in $V_5$, we can define a rational map from $\Gamma_1 \times \Gamma_2$ (which, being a product of rational varieties, is itself a rational variety) to $V_5$ using the method of secants. Namely, given a point on each of the lines, we construct the secant through them and try to find its intersection with the cubic. This construction may fail either because the lines themselves intersect (which can happen at only one point in $\Gamma_1 \times \Gamma_2$) or because the secant is itself a line in $V_5$ (which can happen only at a finite number of points in $\Gamma_1 \times \Gamma_2$, because there are only a finite number of lines). If it does not fail, we are guaranteed to get a rational point in $V_5$, because a non-vanishing cubic with two rational points has a third rational point (possibly repeated). The map is thus defined on a open subset of $\Gamma_1 \times \Gamma_2$ (in the Zariski topology) so is a rational map. When the lines $\Gamma_{1,2}$ are, in addition, disjoint, this map is a birational equivalence. Indeed, it surjects onto an open subset of $V_5$ because every point in $\Q P^3$ (which contains $V_5$) lies on a secant between $\Gamma_1 \times \Gamma_2$ when the two lines are disjoint, and we have removed only a finite number of points from the domain of the map. Moreover, it must inject. For if two distinct secants between the two lines were to  intersect $V_5$ at the same point, we could use them to construct a plane which would, perforce, also contain the two lines. But two lines in a plane must intersect somewhere in projective space, contradicting the hypothesis that they are disjoint.

      We can repeat this argument verbatim over $\R$ (or $\C$), using the same rational lines. The map that results (which is defined by rational functions) is manifestly  continuous in the topology induced by the usual euclidean topology on $\R$ (not merely the Zariski topology). Since the rational points in $\Gamma_1 \times \Gamma_2$ are dense in the real points and since the domain of the map is dense (in this topology) over the reals in $\Gamma_1 \times \Gamma_2$ (with the euclidean topology), it follows that the rational points on $V_5$ are dense (in this topology) in the real points on $V_5$. The same will be true if we intersect $V_5$ with any {\em open} set (in this topology). So the ordered rational points on $V_5$ are dense in the ordered real points, as are the orderable points, as are the nonpalindromic points. All of this can be seen in \cref{fig:figclebschscanv7}. The Figure also shows that the unorderable rational points (of which only 16 out of 25 are visible in this affine patch) are very far from being dense in the unorderable real points (which form a union of one-dimensional manifolds). This turns out to be a feature rather than a bug, since it makes our algorithm for finding anomaly-free irreps far more efficient than one might na\"{\i}vely expect.
\subsection{Parameterization}
We now discuss how one can explicitly generate chiral anomaly-free irreps of $\mathfrak{su}_5$, {\em e.g.} for the purposes of doing phenomenology. As we have seen, (the equivalence classes of) these correspond to nonpalindromic ordered rational points on $V_5$.

The arguments based on the method of secants in the previous Subsection suggest a way to proceed. Namely, if we choose any disjoint pair of lines $\Gamma_{1,2}$ in $V_5$, then we can construct a bijective map from a set containing all of $\Gamma_1 \times \Gamma_2$ apart from a finite number of points to a dense open subset of $V_5$. The image of this map misses at most points lying on rational lines in $V_5$. It is easy to check that the only rational line in the ordered region on $V_5$ is the palindromic line. It follows that by picking any pair of disjoint lines and using the method of secants in this way, we can find all nonpalindromic ordered rational points and so all chiral anomaly-free representations.

The problem, of course, is that the image of this map is dense in $V_5$, while the ordered region is not. It thus contains many rational points that are not in the ordered region. Indeed, since the orderable region consists of a disjoint union of $5!/2 = 60$ copies of the ordered region and is dense in $V_5$, we expect, {\em ceteris paribus}, that fewer than one out of every sixty points that we might generate by means of a scan will lie in the ordered region and so correspond to irreps. Such a method is rather inefficient (though still much better than using trial-and-error, as in Ref. \cite{Eichten_1982}).

One way to try to improve things is to restrict the domain of the map to the preimage of the nonpalindromic ordered region, so as to obtain a bijective parameterization. Doing so requires us to solve analytically for the preimage of the boundary of the weakly-ordered region. This can be done for $n=5$,\footnote{Doing so gives an independent proof that the ordered region is diffeomorphic to $\R^2$ for $n=5$.} but we were unable to generalize it to arbitary $n$.

A method which does easily generalize is to exploit the $S_5$ symmetry of $V_5$. To wit, suppose the rational point spat out by a map as above is not in the ordered region. Provided it is in the order{\em able} region, {\em i.e.} in one of the 59 copies of the ordered region obtained under the action of $S_5$, we can simply permute its co-ordinates $\sigma_i$ until it is in the ordered region. Since the orderable region is dense in $V_5$, we now obtain a parameterization that works almost all of the time.

A further advantage of this algorithm is that it does not matter which pair of disjoint lines one chooses to implement it. Indeed, the action of $S_5$ on $V_5$ induces an action on the set of pairs of disjoint lines in $V_5$. An explicit computation shows that there are $15\times4=60$ such pairs, and any one pair is stablized by a $\Z/2$ subgroup of $S_5$. It follows by the orbit-stabilizer theorem that the induced action is transitive. Thus, when using a disjoint pair of lines to find points in the orderable region (which is fixed by $S_5$), it makes no difference which pair we choose.

In fact, this method works far better than expected, because the unorderable rational points are {\em not} dense in the unorderable real region. Whereas the latter is a union of 1-d manifolds (so contains infinitely many points), the former contains only 25 points. To see this, we need to dive briefly into the theory of elliptic curves.
\subsection{The unorderable region via elliptic curves}
The unorderable region is obtained by setting two (or more) of the five $\sigma_i$ equal to each other. This means that it is a union of ${}^5C_2=10$ subregions on the Clebsch diagonal cubic surface, defined by the $S_5$ orbit of
\begin{equation}
	2\sigma_1^3+\sigma_3^3+\sigma_4^3-(2\sigma_1+\sigma_3+\sigma_4)^3=0, \label{eqn:elliptic}
\end{equation}
where $\sigma_2=\sigma_1$ and $\sigma_5=-(2\sigma_1+\sigma_3+\sigma_4)$. Note that the part of this subregion with $\sigma_1=\sigma_2\geq\sigma_3\geq\sigma_4\geq\sigma_5$ belongs to the boundary of the weakly-ordered region on $V_5$. Moreover, we know that there is at least one rational point on this curve, namely the palindromic one, $[\sigma_1:\sigma_2:\sigma_3:\sigma_4:\sigma_5]=[1:1:0:-1:-1]$. We thus have a smooth projective cubic in $\Bbbk P^2$ equipped with a rational point, which defines an \emph{elliptic curve} $C$. It follows that the set $C(\Bbbk)$ of $\Bbbk$-rational points can be endowed with the structure of a group, the \emph{Mordell-Weil group}, with the group multiplication defined via the secant construction. For $\Bbbk=\R$, $C(\R)$ is isomorphic to the (Lie) group $U(1)\times\Z/2$, because the cubic form in \cref{eqn:elliptic} has positive determinant \cite{Silverman_1992}. So each of the ten elliptic curves over $\R$ in the $S_5$-orbit making up the real unorderable region has two connected components, each containing infinitely many points. On the other hand, the methods of Ref. \cite{Silverman_1992} can be used to show that $C(\Q)\cong\Z/6$.
Explicitly, we have
  \begin{multline}
    C(\Q)=\{[0:0:1:0:-1], [1:1:-1:-1:0], [0:0:0:1:-1], \\ [1:1:-1:0:-1], [0:0:1:-1:0], [1:1:0:-1:-1]\},
  \end{multline}
where the elements as ordered as the powers of a generator, beginning with the identity. Note that each of the first, third and fifth points is where three of the ten real elliptic curves meet (for example, $[0:0:1:-1:0]$ lies on the curves with $\sigma_1=\sigma_2$, $\sigma_2=\sigma_5$ and $\sigma_1=\sigma_5$), while the other three points each lie on the intersection of only two ($[1:1:0:-1:-1]$ satisfies $\sigma_1=\sigma_2$ and $\sigma_4=\sigma_5$ simultaneously, for example). Thus, there are a total of only $3\times10/3+3\times10/2=10+15=25$ unorderable rational points on the Clebsch diagonal cubic surface, given by the $S_5$ orbits of the points in $C(\Q)$. What is even more interesting is that each of the ten unorderable rational points lying on three of the ten elliptic curves also simultaneously lies on three of the fifteen rational lines. They will be familiar to geometers as the ten Eckardt points of the Clebsch diagonal cubic surface.

\begin{figure}
	\centering
	\includegraphics[width=\linewidth]{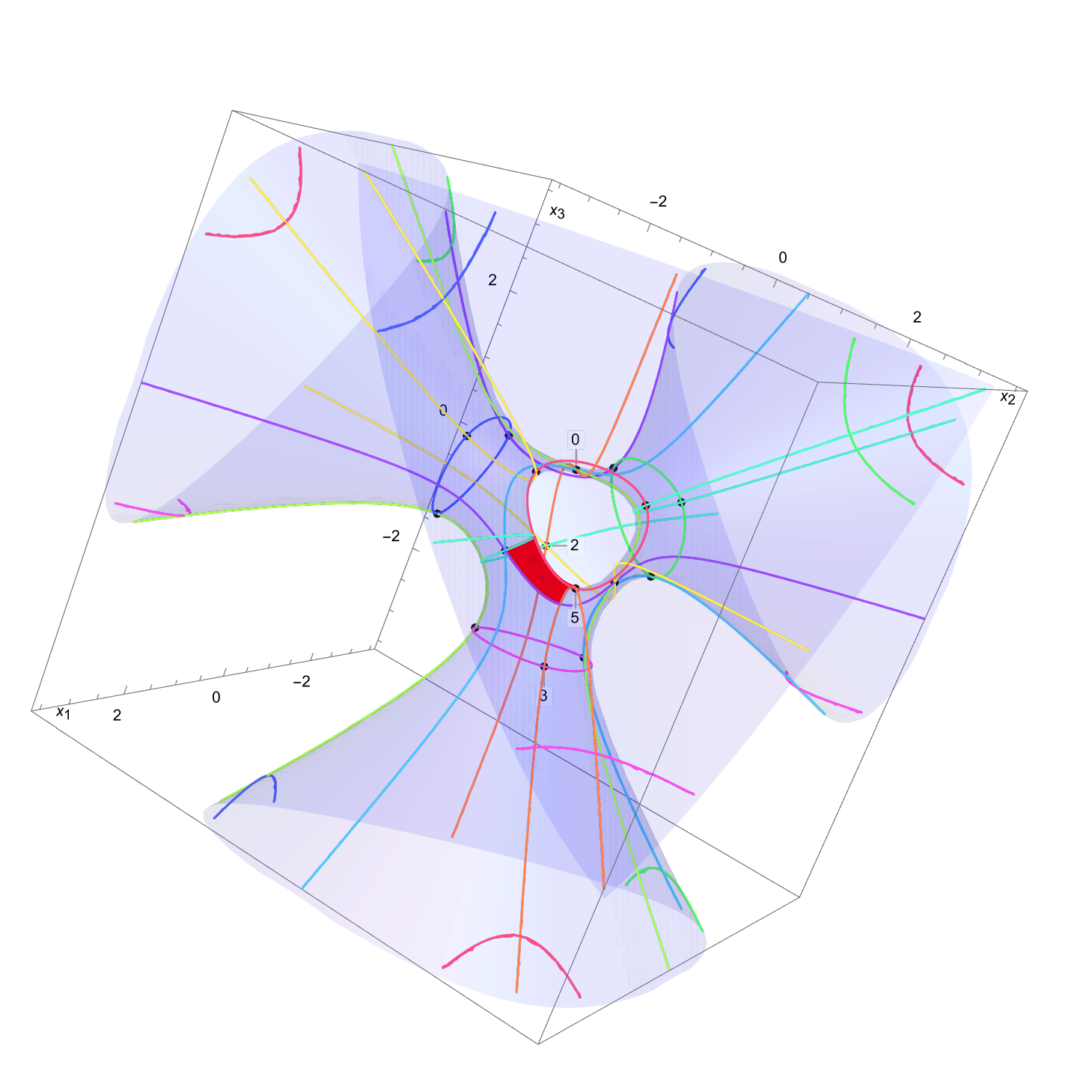}
	\caption{Elliptic curves on the Clebsch diagonal cubic surface corresponding to the unorderable region for $\mathfrak{su}_5$. There are 10 real curves (each shown in a different colour). The orange curve is the curve $C(\R)$ discussed in the main text. Each curve contains 6 rational points and collectively they contain 25 points, of which 16 are visible (in black) in this affine patch. Four of the points in $C(\Q) \cong \Z/6$ discussed in the main text are visible, labelled as powers of the generator (the generator itself is at infinity).}
	\label{fig:figclebschellipticcurvesv6}
\end{figure}

\cref{fig:figclebschellipticcurvesv6} illustrates these results. Of the 16 rational unorderable points that are visible in this affine patch, 
12 correspond to double intersections of the real elliptic curves while the remaining 4 correspond to triple intersections. The curve $C(\R)$ is shown in orange.

\subsection{Comparison with Eichten {\em et al.} \cite{Eichten_1982}}
To illustrate the power of our approach, let us make a brief comparison with the trial-and-error scan of 
Eichten {\em et al.} \cite{Eichten_1982}. As well as being vastly more efficient, the insight that the geometric picture gives allows us to see at a glance that the results claimed in Ref. \cite{Eichten_1982} cannot be correct.

Indeed, it is claimed there that the chiral anomaly-free irreps of $\mathfrak{su}_5$ with dimension up to $4\times 10^9$ have $(q_1,q_2,q_3,q_4)$ given by $(1,8,4,4), (2,9,2,6)$ and $(8,8,16,2)$. From our point of view, the appearance of $(8,8,16,2)$ is hardly surprising, since it is simply the dual of the chiral anomaly-free irrep obtained by doubling $(1,8,4,4)$. Similar logic applied to $(2,9,2,6)$ shows that Ref. \cite{Eichten_1982} missed the irrep $(4,18,4,12)$ (of dimension $3\,121\,637\,376$). By using the method of secants as described above, we find two more irreps missed by Ref. \cite{Eichten_1982}, namely $(1,18,13,6)$ (of dimension $2\,454\,589\,176$) and $(9,17,1,16)$ (of dimension $2\,669\,468\,724$).
\subsection{Topology of the ordered region}
\begin{figure}
	\centering
	\includegraphics[width=\linewidth]{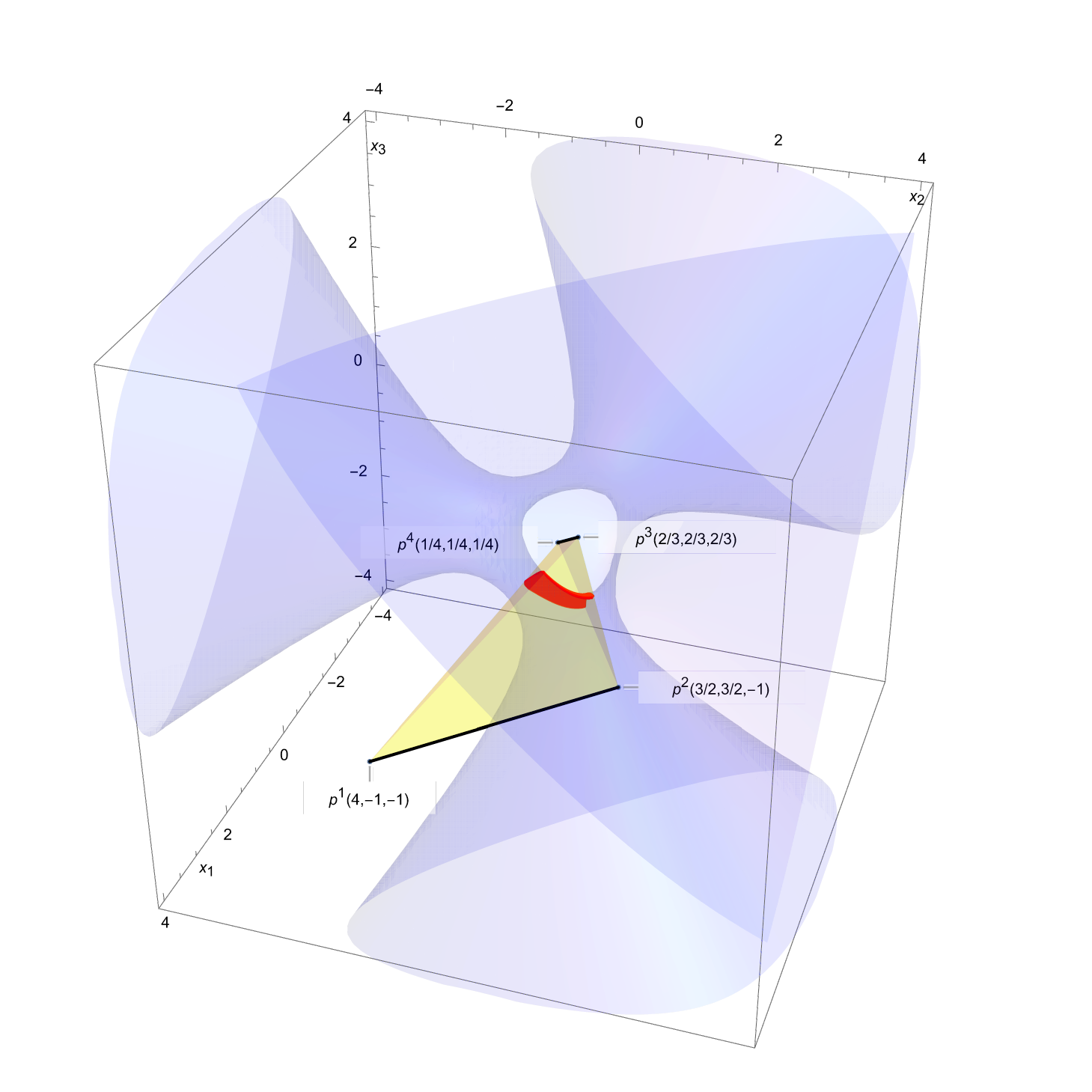}
	\caption{An affine patch of the Clebsch diagonal cubic surface (in blue). The weakly ordered region on $\R P^3$ is shown as the yellow tetrahedron. Secants between the two edges 
shown in black intersect the surface exactly once and define a homeomorphism between the weakly-ordered region on the surface (in red) and a square. This descends to a diffeomorphism between the ordered region on the surface (corresponding to anomaly-free irreps of $\mathfrak{su}_5$) and $\R^2$. 
	\label{fig:figorderedregionv2}}
\end{figure}

We now wish to show that the ordered region on $V_5$ over the reals is diffeomorphic to $\R^2$. To do so, we first observe that the weakly-ordered region on $\R P^{3}$ is the tetrahedron shown in \cref{fig:figorderedregionv2}, on whose vertices the five $\sigma_i$ assume only two distinct values. In the coordinates $x_{1,2,3}$ on the affine patch, these are the points $p^1(4,-1,-1)$, $p^2(3/2,3/2,-1)$, $p^3(2/3,2/3,2/3)$ and $p^4(1/4,1/4,1/4)$.

Next, we observe that each secant between the edges $\overline{p^1p^2}$ and $\overline{p^3p^4}$ is contained in the tetrahedron, that every point in the tetrahedron lies on such a secant, and that the weakly-ordered region on $V_5$ is contained in the tetrahedron as well.

Now, it is not hard to show that $\sum_{i=1}^5\sigma_i^3$ (which defines $V_5$) is everywhere positive on $\overline{p^1p^2}$ and everywhere negative on $\overline{p^3p^4}$. Each secant between these two edges must therefore intersect $V_5$ at least once. If we can show that it is exactly once (so that we get a continuous map), then the observations in the previous paragraph will allow us to establish a homeomorphism between the product of the two edges ({\em i.e.} a square) and the weakly-ordered region on $V_5$. So we are left with showing that each secant intersects $V_5$ exactly (or even just no more than) once. 
Now, on any such secant, the expression $\sum_{i=1}^5\sigma_i^3$ can be written as a cubic\footnote{{\em A priori}, it is possible that the polynomial is of lower order than cubic, but the argument that follows will cover such degenerate cases.} polynomial in a single variable, say $t \in [0,1]$, that changes sign from positive at $t=0$ (the endpoint on $\overline{p^1p^2}$) to negative at $t=1$ (the one on $\overline{p^3p^4}$). When the coefficient of the cubic term in this polynomial is non-negative, we of course have exactly one root in $[0,1]$. When the coefficient is negative, bad things could happen, but an explicit computation (which we omit) shows that the polynomial is then always convex in $[0,1]$, and this is enough for us to conclude that we have exactly one root in $[0,1]$. Again, the geometric picture can be seen in \cref{fig:figorderedregionv2} (which is perhaps already proof enough for many).

By restricting to the interior, we obtain our desired result that the region of ordered real points on $V_5$ is diffeomorphic to $\mathbb{R}^2$. 

\section{The case of $\mathfrak{su}_n$}
We now wish to generalize our results for $n=5$ to higher $n$. In particular, we will prove the following theorems:
\begin{enumerate}
	\item The variety $V_n$ is rational over $\Q$, $\R$, or $\C$ (so we can parameterize the rational points);
	\item The orderable rational points are dense (in the Zariski topology) in $V_n$ over $\Q$ (so we can develop an efficient algorithm to find the ordered rational points, and hence the anomaly-free irreps);
	\item The ordered rational points are dense (in the euclidean topology) in the ordered real points, which are in turn diffeomorphic to $\R^{n-3}$;
	\item The palindromic ordered rational points are dense (in the euclidean topology) in a submanifold of $\R^{n-3}$ diffeomorphic to 
	$\R^{(n-3)/2} $ for $n$ odd or $\R^{(n-2)/2}$ for $n$ even (so the chiral anomaly-free irreps, if $n \geq 5$ such that they exist at all, overwhelm the non-chiral ones). 
\end{enumerate}

The last theorem hints that there are significant differences between the cases where $n$ is odd {\em versus} where it is even, and indeed this is the case. Most strikingly, $V_n$ is a smooth variety for all odd $n$, but has $\frac{n!}{2(n/2)!^2}$ singular points for all even $n$, given by the $S_n$-orbit of $[\sigma_1:\dots:\sigma_{n/2}:\sigma_{n/2+1}:\dots\sigma_n] = [+1:\dots:+1:-1:\dots:-1]$.

For the purposes of proving that $V_n$ is rational and constructing an algorithm for finding explicit anomaly-free irreps, the singular nature of $V_n$ for $n$ even is very much a good thing. Indeed, along with the methods based on constructing secants between two points on the surface, we also have the alternative of constructing lines emanating from one of the singular points. This leads to somewhat simpler proofs. For example, let us show that $V_n$ is rational. The singular points of $V_n$ are double points and so any line emanating from it either lies wholly in $V_n$ or intersects the cubic in a third point (which may coincide with the singular point). Since every other point in projective space (ergo every point in $V_n$) can be reached by exactly one line from this singular point, it follows that that the construction allows us to define a rational map from $\Bbbk P^{n-3}$ to $V_n$ that is one half of a birational equivalence.

Nevertheless, since the method of secants works for all $n$, we shall stick with it in what follows. The differences that arise for even $n$ are rather more troublesome when it comes to proving the third theorem in the above list. Indeed, this theorem follows from an intermediate lemma about the shape of the weakly-ordered region on $V_n$ over the reals. For odd $n$, we obtain a direct generalization of the result for $n=5$. Namely, the weakly-ordered region on $V_n$ is homeomorphic to the product of two $(n-3)/2$-simplices. But for even $n$, it is homeomorphic to a space that is the product of an $(n-4)/2$-simplex and an $(n-2)/2$-simplex, but with the $(n-4)/2$-subsimplex at one vertex of the $(n-2)/2$-simplex collapsed to a point (so for $n=6$, for example, we obtain a pyramid on a square base). Happily its interior is still diffeomorphic to $\R^{n-3}$, so all comes out in the wash.
\subsection{Proof of Theorems 1 and 2}
We show that the variety $V_n$ (over $\Bbbk \in \{\Q,\R,\C\}$) is rational by generalizing the method of secants described for $n=5$. Define a $d$-plane $\Gamma:=\sum_{i=1}^{d+1}\alpha_ip_i$, with $d<n-1$, to be a $d$-dimensional projective subspace of $\Bbbk P^{n-1}$, with $[\alpha_1:\dots:\alpha_{d+1}]\in \Bbbk P^d$ parameterising the $d$-plane and $p^i\in \Bbbk P^{n-1}$. A theorem in Ref. \cite{Allanach_2020} then allows us to replace the pair of disjoint rational lines in the $n=5$ case with a pair of disjoint $d$-planes on $V_n$ of dimensions $d_1=d_2=m_o:=(n-3)/2$ if $n$ is odd and $d_1=m_e:=(n-2)/2$, $d_2=m_e-1$ if $n$ is even, such that the method of secants, including the step of permuting coordinates to get ordered points from orderable ones, works as before. Over $\Q$, $V_n$ has planes of the required dimensions; they are described by equations similar to that of the 15 rational lines on the Clebsch diagonal cubic surface in \cref{eqn:rational_lines}:
\begin{align}
	\sigma_{i_1}=\sigma_{i_2}+\sigma_{i_3}=\dots=\sigma_{i_{n-1}}+\sigma_{i_n}=0,&&n\text{ odd},\nonumber\\
	\sigma_{i_1}+\sigma_{i_2}=\sigma_{i_3}+\sigma_{i_4}=\dots=\sigma_{i_{n-1}}+\sigma_{i_n}=0,&&n\text{ even}, \label{eqn:d_planes}
\end{align}
where $\{i_1,\dots,i_n\}=\{1,\dots,n\}$. To get an $(m_e-1)$ plane for even $n$, we can set any $\sigma_{i_j}$ in the second equation to zero (which also sets either $\sigma_{i_{j+1}}$ or $\sigma_{i_{j-1}}$ to zero). An example of a choice of disjoint hyperplanes of the right dimension for each $n$ is given in Ref. \cite{Allanach_2020}. The ancillary \texttt{Mathematica} notebook implements the method of secants to find anomaly-free irreps for $n=5,6,7,8$.

Thus the proof in \cref{sec:rational} generalizes for $n\geq5$: we always have a rational variety, where the birational map is given by the method of secants construction as described.
\subsection{Proof of Theorem 3}
Since the method of secants, which gives a manifestly continuous map in the euclidean topology on $\R$, works for our cubic hypersurfaces with arbitrary $n$, it follows that on all of these hypersurfaces, the (ordered) rational points are dense in the (ordered) real points in the euclidean topology. Recall that in the $n=5$ case, the weakly ordered region on $\R P^3$ is a tetrahedron. For other $n$  $(\geq 2)$, we instead find an $(n-2)$-simplex $\Delta$ on $\R P^{n-2}$, described in the co-ordinates $x_i$ given in \cref{eq:xi} by
\begin{equation}
	x_1\geq x_2\geq\dots\geq x_{n-2}\geq1-x_1-\dots-x_{n-2}\geq-1,\label{eqn:simplex_inequality}
\end{equation}
whose $n-1$ vertices are at (in co-ordinates $(x_1,\dots,x_{n-2})$),
\begin{align}
	p^1&=(n-1,-1,-1,\dots,-1,-1):&&\sigma_2=\sigma_3=\dots=\sigma_n,\nonumber\\
	p^2&=\left(\frac{n-2}{2},\frac{n-2}{2},-1,-1,\dots,-1,-1\right):&&\sigma_1=\sigma_2,\sigma_3=\dots=\sigma_n,\nonumber\\
	p^3&=\left(\frac{n-3}{3},\frac{n-3}{3},\frac{n-3}{3},-1,\dots,-1,-1\right):&&\sigma_1=\sigma_2=\sigma_3,\sigma_4=\dots=\sigma_n,\nonumber\\
	\dots\nonumber\\
	p^{n-3}&=\left(\frac{3}{n-3},\frac{3}{n-3},\dots,\frac{3}{n-3},-1\right):&&\sigma_1=\dots=\sigma_{n-3},\sigma_{n-2}=\sigma_{n-1}=\sigma_n,\nonumber\\
	p^{n-2}&=\left(\frac{2}{n-2},\frac{2}{n-2},\dots,\frac{2}{n-2},\frac{2}{n-2}\right):&&\sigma_1=\sigma_2=\dots=\sigma_{n-2},\sigma_{n-1}=\sigma_n,\nonumber\\
	p^{n-1}&=\left(\frac{1}{n-1},\frac{1}{n-1},\dots,\frac{1}{n-1},\frac{1}{n-1}\right):&&\sigma_1=\dots=\sigma_{n-1},
\end{align}
where we also give the corresponding conditions on the homogeneous coordinates $\sigma_i$. 

On this affine patch, defined by $\sigma_n \neq 0$, we are free to set $\sigma_n = -1$; doing so, the polynomial
\begin{equation}
	\sum_{i=1}^{n-1}\sigma_i^3-\left(\sum_{i=1}^{n-1}\sigma_i\right)^3
\end{equation}
defining the cubic hypersurface becomes
\begin{equation}
g(x):=\sum_{i=1}^{n-2}x_i^3+\left[1-\left(\sum_{i=1}^{n-2}x_i\right)\right]^3-1.
\end{equation}

The next step in the proof for $n=5$ was to exploit the fact that there exist two edges of the tetrahedron that are mutually disjoint and that do not intersect the cubic hypersurface. Unfortunately, this does not generalize to even $n$, where we find that the vertex $p^{n/2}=(1,1,\dots,-1,-1)$ is a singular point on the variety. So we proceed by considering the different parities of $n$ separately.
\subsubsection{Odd $n$}
For odd $n:=2k-1$, the set of vertices $\{p^1,\dots,p^{2k-2}\}$ of $\Delta$ can be partitioned into two subsets. The first subset, $\{p^1,\dots,p^{k-1}\}$ contains the vertices of a $(k-2)$-subsimplex $\delta^+$, whose points have $\sigma_k=\dots=\sigma_{2k-1}<0$. The other subset, $\{p^k,\dots,p^{2k-2}\}$, contains the vertices of another $(k-2)$-subsimplex $\delta^-$, whose points have $\sigma_1=\dots=\sigma_k>0$. Note that $\delta^+$ and $\delta^-$ are disjoint. Applying Jensen's inequality, one can show that $g(x)$ is positive on $\delta^+$ and negative on $\delta^-$. We can now construct secants connecting a point on $\delta^+$ to a point on $\delta^-$ as before, and the proof in the $n=5$ case that the associated cubic polynomial in $t\in[0,1]$ has only one zero in this interval carries over to higher (odd) $n$ as well. In this way we construct a homeomorphism between $\delta^+\times\delta^-$, which is the Cartesian product of two simplices of dimension $k-2=(n-3)/2$, and the weakly ordered real region on $V_n$. Its interior, the ordered real region on $V_n$, is diffeomorphic to $\R^{n-3}$.
\subsubsection{Even $n$}
For even $n:=2k$, we construct instead the $(k-2)$-subsimplex $\delta^+$, whose vertices are in the set $\{p^1,\dots,p^{k-1}\}$, on which $g(x)$ is positive, and the $(k-1)$-subsimplex $\delta^-$, with vertex set $\{p^k,\dots,p^{2k-1}\}$, on which $g(x)$ is negative except at the double point $p^k$ where $g(x)$ vanishes. The construct of the previous paragraph and the associated proofs still follow through for all line segments from $\delta^+$ to $\delta^-$ that do not end on $p^k$. For line segments that do end on $p^k$, it turns out that this is their only point of intersection with the weakly ordered region on $V_n$.\footnote{To prove this, construct the auxiliary subsimplex $\delta'\supset\delta^+$ with vertex set $\{p^1,\dots,p^k\}$, and observe that (i) $g(x)$ is positive everywhere on this subsimplex except at $p^k$, where it vanishes, and (ii) every such line segment lies entirely on this subsimplex.} The effect of this is that this region is homeomorphic to the Cartesian product of a simplex $\delta^+$ of dimension $k-2=(n-4)/2$ and another, $\delta^-$, of dimension $k-1=(n-2)/2$, except that the copy of $\delta^+$ at one vertex of $\delta^-$ (the double point) is collapsed to a single point. Nevertheless, its interior is still diffeomorphic to $\R^{n-3}$.
\subsection{Proof of Theorem 4}
Recalling that the palindromic points are defined by $[\sigma_1:\dots:\sigma_n]=[-\sigma_n:\dots:-\sigma_1]$, it is trivial to check using \cref{eqn:d_planes} that the ones with rational coordinates are dense (in the euclidean topology) on a $d$-plane on dimension $m_o=(n-3)/2$ for odd $n$ or $m_e=(n-2)/2$ for even $n$.
\section{Reducible representations}
For a small number of theories with nonsemisimple gauge groups, explicit representations that are free of local anomalies have been found on a case-by-case basis \cite{Allanach_2020_2, Allanach_2020_3, Dobrescu_2020, Abel_2022, Costa_2020, Costa_2019, Rathsman_2019, Lu_2019, Lohitsiri_2020}. However, for a general representation of a general Lie algebra $\mathfrak{g}$, the anomaly cancellation conditions are well known to be complicated, since we can have both abelian and mixed anomalies in addition to those for $\mathfrak{su}_n$ summands that we have considered so far.  Nevertheless, it is still true that these conditions  can be expressed, via the theorem of the highest weight, in terms of polynomial equations taking values in the integers (along with further restrictions). Correspondingly, one can at the very least define an associated affine variety and try to relate the anomaly-free representations to its properties, though we expect that this will in general be difficult. Let us then examine instead just the case where $\mathfrak{g}$ is semisimple.

When $\rho$ is an irrep of $\mathfrak{su}_n$, both its dimension $D(\rho)$ and the associated anomaly $A(\rho)$ can be expressed as homogeneous polynomials (in either the variables $q_i$ or $\sigma_i$) of degrees $n(n-1)/2$ and $n(n-1)/2+3$ respectively. In fact, much more is true, in that $D(\rho)$ is a factor of $A(\rho)$. Explicitly, we have \cite{Georgi_1976, Okubo_1977}
\begin{alignat}{2}
	D(\rho)&=\prod_{j=1}^{n-1}\left[\frac{1}{j!}\prod_{k=j}^{n-1}\left(\sum_{i=k-j+1}^kq_i\right)\right]&&=\frac{\prod_{j<k}^n(\sigma_j-\sigma_k)}{n^{n(n-1)/2}\prod_{i=1}^{n-1}i!},\\
	\frac{A(\rho)}{D(\rho)}&=\frac{2(n-3)!}{(n+2)!}\sum_{i,j,k=1}^{n-1}a_{ijk}q_iq_jq_k&&=\frac{2}{n^2(n^2-1)(n^2-4)}\sum_{i=1}^n(\sigma_i)^3,
\end{alignat}
where $a_{ijk}$ is described in \cref{eqn:aijk}, and we remind the poor reader who has struggled this far that our $\sigma_i$ are $n$ times those of Okubo. From this, we conclude that given an arbitrary representation of $\mathfrak{su}_n$, by reducing it and using the fact that the anomaly of the sum is the sum of the anomalies, we get that its anomaly is once again a homogeneous polynomial, but now of degree $n(n-1)/2+3$ (in say the integers $q_i$ or $\sigma_i$). So it makes sense to put the representations into equivalence classes under an overall scaling by a positive integer and just as in the case of irreps, we see that given just one anomaly-free chiral $(m-1)$-times reducible representation we can easily find infinitely many more. Solving the anomaly cancellation equation, however, would require finding all integer solutions to a polynomial equation of degree $n(n-1)/2+3$ in $m(n-1)$ variables (each irrep factor being characterized by $n-1$ of them). For a representation that is not arbitrary, but rather is a product of irreps, we can say much more, but we will leave that story for another night.

\acknowledgments{We thank Robert Bourne, Alex Colling, Jun Liu and Timothy Moy for discussions.\linebreak This work has been partially supported by STFC consolidated grants ST/T000694/1 and ST/X000664/1 and a Trinity-Henry Barlow Scholarship.}
\bibliography{varieties_references_submission}    
\end{document}